\documentclass[twocolumn,aps,superscriptaddress]{revtex4-1}
\usepackage{palatino}
\usepackage[latin1]{inputenc}
\usepackage{epsf}
\usepackage{amsmath,amssymb}
\usepackage{latexsym}
\usepackage{calc}

\usepackage{float}

\usepackage[usenames,dvipsnames]{xcolor}

\usepackage{graphicx}
\usepackage{hyperref}
\newcommand{\ben}{\begin{equation}}
\newcommand{\een}{\end{equation}}
\newcommand{\bea}{\begin{eqnarray}}
\newcommand{\eea}{\end{eqnarray}}

\def\sss{\scriptscriptstyle\rm}

\def\1s{_{1,\sss S}}
\def\2s{_{2,\sss S}}
\def\x{_{\sss X}}
\def\c{_{\sss C}}
\def\s{_{\sss S}}
\def\xc{_{\sss XC}}

\def\H{_{\sss H}}

\def\ext{_{\rm ext}}

\def\br{{\bf r}}

\def\bj{{\bf j}}

\begin{document}
 \title{ The Exact Exchange-Correlation Potential in Time-Dependent Density Functional Theory: Choreographing Electrons with Steps and Peaks}
 \author{Davood Dar}
 \affiliation{Department of Physics, Rutgers University, Newark 07102, New Jersey USA}
 \author{Lionel Lacombe}
 \affiliation{Laboratoire des Solides Irradi{\'e}s, {\'E}cole Polytechnique, Institut Polytechnique de Paris, F-91128 Palaiseau, France}
 \author{Neepa T. Maitra}
 \affiliation{Department of Physics, Rutgers University, Newark 07102, New Jersey USA}
 \email{Corresponding Author: neepa.maitra@rutgers.edu}
 \date{\today}



\begin{abstract}
The time-dependent exchange-correlation potential has the unusual task of directing fictitious non-interacting electrons to move with exactly the same probability density as true interacting electrons. This has intriguing implications for its structure, especially in the non-perturbative regime, leading to step and peak features that cannot be captured by bootstrapping any ground-state functional approximation. We review what has been learned about these features in the exact exchange-correlation potential of time-dependent density functional theory in the past decade or so, and implications for the performance of simulations when electrons are driven far from any ground-state. 
\end{abstract}

\maketitle
\section{Introduction}
\label{sec:intro}
Time-resolved dynamics of electrons in molecules and solids have become increasingly relevant over the past decades. A description beyond equilibrium electronic structure and excitation spectra is necessary in many applications of fundamental and technological importance: photovoltaic processes, photocatalysis, radiation damage in biomolecules, nanoscale conductance devices, time-resolved pump-probe spectroscopies, and strong light-matter coupling for quantum-based technologies. 
Scalable theoretical methods to computationally model coupled electron, ion, and photon dynamics help in interpreting and predicting experiments, and to suggest new systems with improved functionalities. Arguably, the electron-component of the problem is the most challenging: there are many of them, they interact with each other as well as with the nuclei and light fields, and they require a quantum mechanical description, unlike nuclei and photons for which, for different reasons,  a classical description may suffice.

To this end, time-dependent density functional theory (TDDFT)~\cite{RG84,TDDFTbook12,Carstenbook,M16} has emerged as a method of choice. 
TDDFT is inherently a ``real-time" method in that dynamics is woven into the formulation from the very beginning with the foundational theorem developed for general time-dependent evolution of arbitrary initial states. While perturbations around the ground-state yield a formalism for excited state energies, their couplings, and response, it is particularly the generality and practical efficiency of the real-time formulation that has led to the possibility of applications on large systems which could not be done otherwise, e.g. ~\cite{DAGBSC17,SZYYK21,JKC22,LGIDL20,S21,MFHCGLS22, DVCSR22,SK18, YNNY18, BCV22, Floss18,EMDSG16,UAC18,RFSR13}.

 In a sense, TDDFT is an extension of its ground-state counterpart (DFT)~\cite{HK64,KS65}. 
 The central object is the one-body density, $n(\br,t)$, which is the probability density of finding any one electron at point $\br$ in space at time $t$: $n(\br,t) = N \sum_{\sigma_1...\sigma_N}\int d^3r_2...d^3r_N \vert \Psi(\br \sigma_1, \br_2 \sigma_2...\br_N\sigma_N)\vert^2$ where $\sigma_i$ represents the spin coordinate, and $N$ is the number of electrons in the system. 
 Just like DFT, TDDFT allows one to bypass having to solve the analytically and computationally complicated many-body time-dependent Schr\"odinger equation (TDSE) and instead only requires one to solve a set of time-dependent single-particle equations involving a modified one-body potential whose solutions yield the same time-dependent one-body density as that of the true system. This drastically reduces the amount of computational effort in obtaining quantum many-body dynamics. 
 
This is an audacious concept: the idea that a set of non-interacting electrons reproduces the density of interacting electrons seems fantastical, especially in the time-dependent case, where the motion of the electrons driven by some external perturbation or by nuclear motion is affected by their mutual repulsion in an intricate dance. The one-body potential somehow directs the non-interacting electrons to evolve with the same one-body density as the interacting electrons. Recent work has shown that this choreography results in dynamical step and peak structures that have a non-local density-dependence in time and in space, and 
 are completely missed by the commonly used ``adiabatic" approximations. 
Such approximations utilize ground-state exchange-correlation (xc) functionals, and although ground-state potentials also may feature steps and peaks, they appear only in particular situations such as static correlation or fractionally charged systems, while the TDDFT steps appear quite generically. 

In this review, we revisit what is known about non-adiabatic features of the exact xc potential in TDDFT, focussing on the dynamical steps and peaks. We begin with a brief reminder of the fundamental theory of TDDFT in Sec.~\ref{sec:formalism}, including a discussion of memory-dependence. Section~\ref{sec:steps} demonstrates the dynamical peaks and steps that arise from memory-dependence, by numerical examples as well as analysis of the equations, identifying a ``kinetic  component" in the xc potential as responsible for these features, and discussing the role of the local accelerations in the system. In Section~\ref{sec:He} we make a case study on the helium atom, showing that these features, which have largely been discussed in one-dimensional (1D) systems, persist just as strongly in three-dimensions, and verifying the relevance of the analyses previously made. We summarize in Section~\ref{sec:concs}.

\section{TDDFT formalism}
\label{sec:formalism}
At the heart of TDDFT lies the Runge-Gross theorem~\cite{RG84,GM12,Carstenbook} which, for time-dependent problems, plays an analogous role to the Hohenberg-Kohn theorem for ground-state problems. Namely, for a fixed particle-particle interaction and statistics, it establishes a one-to-one mapping between the  possibly time-dependent external potential acting on the electrons and the time-dependent density for a given initial state:
\ben
    \Psi(0): n \leftrightarrow v\ext
    \label{eq:1-1}
\een
 As a consequence,  all  physical observables can be expressed as functionals of the density and initial state: for a given $\Psi(0)$, $n$ points to a unique $v\ext$ which points to a unique Hamiltonian, which in theory points to a unique time-dependent wavefunction, from which any observable can be extracted in principle. 
 
  But this can be of little more than theoretical interest unless we know how to obtain the density of the interacting system and the observables of interest. 
As in ground-state DFT, one maps the interacting system to a fictitious non-interacting system, the Kohn-Sham (KS) system, whose orbitals are required to reproduce the density at any instant. 
The KS orbitals evolve in the one-body KS potential, $v\s(\br,t)$, according to a single-particle TDSE:
\ben
   \left( -\frac{\nabla^2}{2} +v\s(\br,t)\right)\phi_i(\br,t)=i\partial_t\phi_i(\br,t)\,.
    \label{KS_equation}
\een
(Atomic units are used throughout). The KS potential is written as the sum of three terms in a similar way to DFT:
\ben
    v\s(\br,t)=v\ext(\br,t)+v\H[n](\br,t)+v\xc[n;\Psi_0,\Phi_0](\br,t)
     \label{Vs_equation}
\een
where $v\H[n](\br,t) = \int \frac{n(\br')}{|\br - \br'| }d^3r'$ is the Hartree potential and $v\xc[n;\Psi_0,\Phi_0](\br,t)$ is the xc potential whose exact functional dependence on the density $n$, true initial state $\Psi_0$ and KS initial state $\Phi_0$, is unknown and needs to be approximated in practise. We note that there is much freedom in selecting the initial KS state $\Phi_0$: it can be any wavefunction that reproduces the density of the true interacting initial state and its first time-derivative~\cite{L99,MB01,EM12,FNRM16}. 

The vast majority of applications of TDDFT are for linear response where a perturbative limit of Eqs.~(\ref{KS_equation})--(\ref{Vs_equation}) yields
a Dyson-like equation, or matrix equations, in the frequency domain, whose solution gives excitation energies and oscillator strengths\cite{C95,C96,PGG96,GPG00}. The success of TDDFT in this regime, with the available approximate xc functionals, is incontrovertible, e.g. Refs~\cite{CH12,AJ13,SBMLJ21}, however not without blemishes for certain classes of excitations, e.g. double-excitations and charge-transfer excitations~\cite{M22}, which can create havoc for the black-box use of the method in coupled electron-ion dynamics~\cite{LKQM06}. 
But the fully non-perturbative regime is of particular interest in TDDFT, due to the much harsher computational scaling of alternative methods~\cite{LGIDL20}.

Armed with a good approximation to the xc potential, solving Eq.~(\ref{KS_equation}) with Eq. (\ref{Vs_equation}) yields a good approximation to the density of the physical interacting system, obtained from non-interacting KS electrons, through $n(\br,t) = \sum_{i=1}^N \vert \phi_i(\br,t)\vert^2$. Observables that are directly related to the density, such as the dipole moment, can be directly extracted, while further approximations would be needed to extract other observables, such as momentum distributions~\cite{WB07},  double-ionization cross-sections~\cite{WB06},  and current-densities, where the rotational part differs in general from that of the KS system~\cite{AV05,SK16,DLFM21}. More often than not, these observables are approximated by those of the KS system~\cite{AHC18,TNY19,GHR17}; an open question is how errors from the xc functional itself compare with the (usually unacknowledged) errors from the observable evaluation. 

 An exact expression for the xc potential can be derived by equating the second time-derivative of the density, $\ddot{n}(\br,t)$, for the KS system with that of the interacting system. While the Heisenberg equation of motion for the density gives the continuity equation $\dot{n} = -\nabla\cdot{\bj}$, $\ddot{n}(\br,t)$ can then be obtained from the equation of motion for the current-density $\bj(\br,t)$. Subtracting the equation obtained for the interacting system from that for the KS system results in the decomposition of the xc potential, $v\xc(\br,t)$ into  kinetic (T) and interaction (W) terms $v\xc(\br,t) =v_c^{\rm T}(\br,t) +v\xc^{\rm W}(\br,t)$, that have the following structure~\cite{L99,LFSEM14,FNRM16,FLNM18,LM18,LM20b} :
\ben
    \nabla\cdot\left( n  \nabla v\xc^{\rm W}\right) = 
  \nabla\cdot\left( n (\br,t)\int n\xc(\br,\br',t)\nabla w(|\br'-\br|)d^3\br'\right)
  \label{eq:vxcW}
\een

\ben
 \nabla\cdot\left( n    \nabla v_c^{\rm T}\right)=\nabla\cdot\left({\cal D}_{\br',\br}\Delta\rho_1(\br',\br,t)|_{\br'=\br}\right) \,,
\label{eq:vcT}
\een
(with the $(\br,t)$-dependences on the left-hand-side understood)
 where $n\xc(\br,\br',t)$ is the xc hole, defined through the diagonal two-body density matrix $\rho_2(\br,\br';\br,\br') = N(N-1)\int dr_3...dr_N \vert \Psi(\br,\br',\br_3..\br_N)\vert^2 = n(\br',t)\left(n(\br,t) + n\xc(\br,\br',t)\right)$ and $\Delta \rho_1(\br',\br,t)= \rho_1(\br',\br,t)-\rho_{1,S}(\br',\br,t)$ is the difference between the spin-summed one-body density matrix of the true interacting system $\rho_1(\br',\br,t)$ and  that of the Kohn-Sham system, $\rho_{1,S}(\br',\br,t)$. The differential operator ${\cal D}_{\br',\br} = \frac{1}{4}(\nabla'-\nabla)(\nabla^2-\nabla'^2)$. 
 Eqs.~(\ref{eq:vxcW})-(\ref{eq:vcT}) give an exact expression for the xc potential in terms of the exact xc hole, and the exact and KS one-body reduced matrices. They are useful for analysis of TDDFT and understanding errors in approximations. Further, they offer a starting point for approximations: although $\rho_{1,S}(\br',\br,t)$ is accessible in a KS evolution, $\rho_1$ and $n\xc$ need to be approximated in terms of KS quantities. We will also return to this in Sec.~\ref{sec:steps}.
 
\subsection{Memory: History and Initial-State Dependence}
\label{sec:memory}
In traditional wavefunction-based quantum mechanics, knowing the wavefunction $\Psi(\br_1...\br_N,t) $ at any time $t$ is enough to know all properties of the system at that time; its value at earlier times is not required. 
The same cannot be said about the one-body density. 
While the TDDFT reformulation of the many-body problem in terms of its one-body time-dependent density $n(\br,t)$ drastically simplifies the problem from the computational viewpoint, it inevitably introduces some complications. In particular, the xc potential is memory-dependent in that the functional $v\xc[n,\Psi_0,\Phi_0](\br, t)$ depends not only on the instantaneous density $n(\br,t)$ but also on the history of the density $n(\br,t'<t)$ and the initial states $\Psi_0, \Phi_0$. 
This follows directly from the Runge-Gross theorem Eq.~(\ref{eq:1-1}): the mapping is between the density- and potential- functions over space and time, and is one-to-one for a given initial state. This would mean $v\ext(\br,t)$ functionally depends on the density over all times  but due to causality it depends only the history of the density, not its future. The initial-state dependence means that the same time-dependent density can be  obtained by propagating in two different potentials if the systems begin in different initial states. Applying Eq.~(\ref{eq:1-1}) to the  KS system yields that $v\s(\br,t)$ functionally depends on the history of the density and the KS initial state, and thus $v\xc(\br,t) = v\s(\br,t) - v\ext(\br,t) -v\H(\br,t)$  functionally depends on both the true and KS states: $v\xc[n; \Psi(0), \Phi(0)](\br, t)$. As mentioned earlier, one can begin in any initial KS state that reproduces the density of the initial interacting state and its first time-derivative; the structure of the exact xc potential has a strong dependence on this choice~\cite{EM12,FNRM16,FLNM18,LM18,SLWM17,NRL13}. 

There is an intimate connection between history-dependence and initial-state-dependence that can be useful. What we take as the  ``initial" time can be re-set, and if we know the wavefunctions at the reset time, then we can evaluate the xc potential on the domain of those wavefunctions and a truncated history, i.e.~\cite{MBW02}
\ben
v\xc[n;\Psi_0,\Phi_0](\br,t) = v\xc[n_{t'};\Psi_{t'},\Phi_{t'}](\br,t), \;\;\; t'\le t
\label{eq:mem_cond}
\een
where $n_{t'}$ means the density function over the domain from time $t'$ to $t$.
A useful consequence  is that if the xc potential is found at some time $t$ for a particular dynamics of a system, then in {\it any} dynamics where the exact same  interacting and KS states happen to be reached at some time, then the xc potential at that time will be the same.  We will exploit this in Section~\ref{sec:He} to emphasize the generality of the features of the xc potential found there.

Almost all calculations today however completely neglect memory. They use an adiabatic approximation, in which the instantaneous density is input into a ground-state approximation: $v\xc^A[n;\Psi_0,\Phi_0](\br,t) = v\xc^{\rm g.s.}[n(t)](\br)$. Such an approximation has two sources of error: one arising from the approximation made for the ground-state functional, and the other from the adiabatic approximation itself. To isolate the error from the adiabatic approximation itself, the adiabatically-exact approximation~\cite{TGK08} is defined as
$ v\xc^{\rm adia-ex}[n;\Psi_0,\Phi_0](\br,t) = v\xc^{\rm exact-g.s.}[n(t)](\br)$
which can be a useful analysis tool in cases where the exact ground-state xc potential can be computed (usually model 1D systems). 
Due to the lack of memory, the adiabatic approximation leads to large errors in some applications, sometimes failing completely~\cite{RN11,RN12,RN12b,RN12c,HTPI14,PI16,WU08,GDRS17,U06}, but in other cases it has been found to yield good predictions~\cite{DAGBSC17,SZYYK21,JKC22,LGIDL20,S21,MFHCGLS22, DVCSR22,SK18, YNNY18, BCV22, Floss18,EMDSG16,UAC18,RFSR13}, even when the system is far from a ground-state. It is not completely understood why: possible reasons include, that the adiabatic approximation satisfies a number of exact conditions that are important in the time-dependent case~\cite{TDDFTbook12},  that in some applications a strong external field dominates over xc effects in driving the dynamics and that partial compensation of self-interaction in the Hartree potential, even at the ground-state level, is enough, especially when the observables involve averaging over the details of the density distribution.  Recently it was further argued that an indicator of the expected success or failure of the adiabatic approximation lies in the natural orbital occupation numbers~\cite{LM20b}:  if the initial KS state has a configuration close to that of the true initial state, and the natural orbital occupation numbers of the true system do not evolve significantly in time,  the adiabatic approximation may make good predictions even for strongly non-perturbative dynamics. 

\subsection{Examples showing the relevance of memory}
Model systems have been crucial in understanding cases where the adiabatic approximation fails, because numerically exact solutions are available and because the exact xc potential can be extracted to compare with approximations~\cite{V08,KPV11,EFRM12,RG12,FM14,FM14b,K16,SLWM17,DSH18,CPRS18}.  Fig.~\ref{fig:1Dexamples} shows the errors that  adiabatic approximations make in  a variety of  studies. 

The top panel shows the electronic dipole for field-free evolution of a state that is a 50:50 superposition of the ground and first-excited singlet state of a 1D soft-Coulomb-interacting He atom~\cite{EFRM12,LFSEM14,FNRM16,FLNM18}; such a state may be reached from a ground state driven by a field that is then turned off, for example. In that scenario, the natural choice for the KS initial state is a Slater determinant, which for our two-electron system, is a doubly-occupied spatial orbital.  As seen in the figure, the adiabatic exact-exchange (AEXX) and local density approximations (ALDA) do not get the period of oscillations correct, and display additional beat frequencies. In Sec.~\ref{sec:steps}, we will see that the exact xc potential has non-adiabatic step and peak features that even the adiabatically-exact approximation lacks. 

Such features have an even more notable effect in electron-scattering, illustrated in panel b in Fig.~\ref{fig:1Dexamples}. This shows the reflection probability of an electron initially in a gaussian wavepacket moving towards a target 1D H atom~\cite{SLWM17,LSWM18}. The approximations severely underestimate the reflection of the wavepacket ($N_R$ is the number of electrons on the right of the atom). The dashed and solid lines represent different choices of initial KS state, both of which have the same density as the true state: In one, a Slater determinant is chosen for example to simulate the return of an ionized electron to its parent, while in the other, the initial KS state is a two-orbital state which has the same configuration as the interacting state with one electron in the bound target ion, and the other in a gaussian wavepacket. Although there are significant differences in the details of the two time-dependent densities ensuing from these states when propagated under ALDA or AEXX~\cite{SLWM17,LSWM18}, with the Slater determinant demonstrating spurious oscillations and reproducing the exact dynamics less accurately initially, neither of them capture the eventual reflection even qualitatively. The exact xc potential shows a step and valley structure that is essential for this effect, and is missing in the adiabatic approximations (see also Sec.~\ref{sec:steps}). This may explain the underestimated predictions of scattering probabilities and energy transfer that has been observed in real systems~\cite{GWWZ14,QSAC17,HKLK09,DLYU17,MUSW17,USW16}.

\begin{figure}
\label{fig:1Dexamples}
\includegraphics[width=0.45\textwidth]{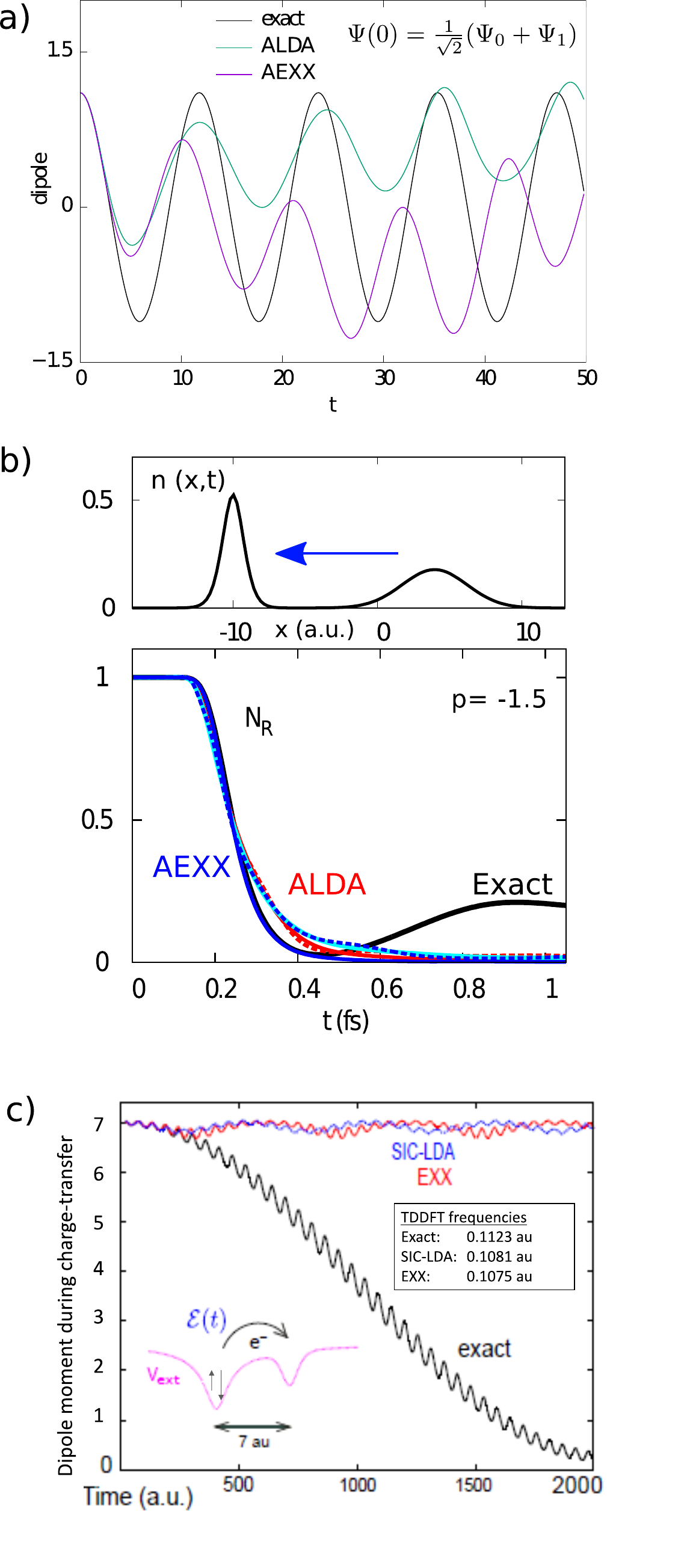} 
\caption{Three  examples illustrating substantial errors in adiabatic TDDFT propagation. a) Dipole moment in field-free propagation of a 50:50 superposition of the ground  and first excited state in a 1D He atom~\cite{EFRM12,LFSEM14,FNRM16,FLNM18}, showing the exact, against AEXX and ALDA. The initial KS orbital is chosen to be a Slater determinant.
b) Scattering of an electron off a 1D H-atom; the top panel shows the density upon approach, while the lower panel shows the integral of the electron density to the right of the atom ($x \ge -5$)~\cite{SLWM17,LSWM18}; dashed curves for the Slater determinant initial state, and solid for a two-orbital state.
c) Resonantly-driven charge-transfer out of the ground-state of the double-well shown. Neither self-interaction-corrected LDA (SIC-LDA) nor AEXX capture the dynamics, despite yielding good approximations for their excitations~\cite{FERM13,M17}. 
}
\label{fig:1Dexamples}
\end{figure}

Panel c of Fig.~\ref{fig:1Dexamples} gives an example of a resonantly-driven charge-transfer out of the ground-state~\cite{FERM13, M17}. A molecule is modeled via an asymmetric double-well, in which the ground-state has two electrons in the left-well. Applying a weak field that is resonant with an excitation to a charge-transfer state triggers a Rabi oscillation, as evident by the large change in the dipole seen as the molecule reaches the charge-transfer state. Although the adiabatic TDDFT approximations shown yield excellent values for the charge-transfer excitation energies, they completely fail to capture the charge-transfer dynamics. Again, dynamical step features develop as the electron transfers (see Sec.~\ref{sec:steps}). 
But adiabatic TDDFT generally fails at resonant driving, even to local excitations~\cite{RB09,EFRM12,RN11,FHTR11}, because it violates the fundamental condition that resonant excitation frequencies of a system should not shift with the instantaneous state; with adiabatic approximations, the density-dependence of the KS potential leads to the response of a non-equilibrium state having spuriously-shifted poles, while the exact generalized xc kernel requires a frequency-dependence to correct this spurious shift~\cite{FLSM15,LFM16}. Although it has not yet been explicitly shown, it is likely this frequency-dependence is related to dynamical steps and peaks in the time-domain. This could lead to some unreliability in TDDFT simulations of pump-probe spectroscopy; absorption peak-shifts have been observed in a number of molecules~\cite{PHI15,PI16,OMKPRV15,GBCWR13}. 



\section{Dynamical steps and peaks}
\label{sec:steps}

\begin{figure}
\includegraphics[width=0.5\textwidth]{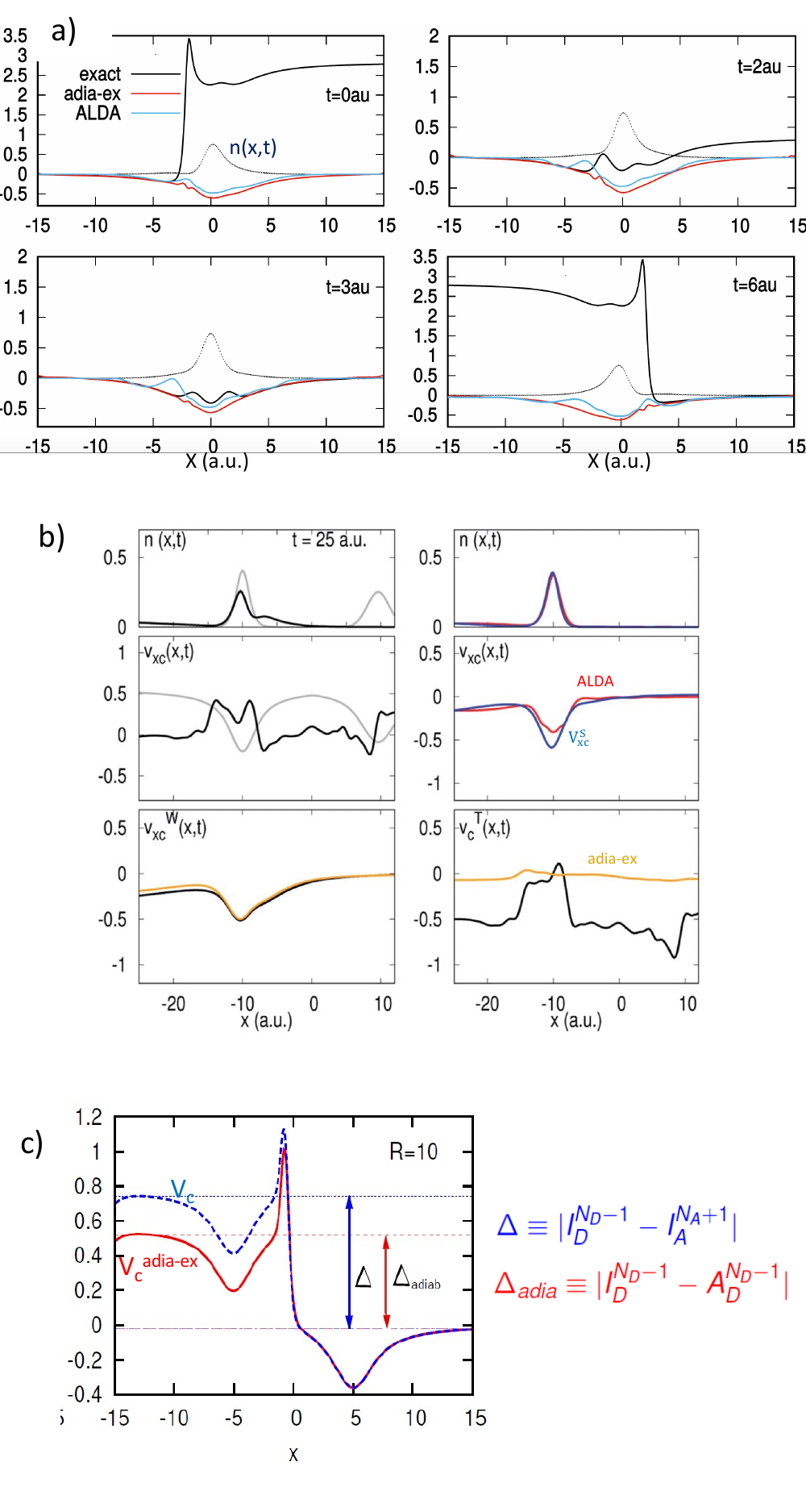} 
\caption{Non-adiabatic step and peak features in $v\xc$ for the examples of Fig.~\ref{fig:1Dexamples}.
(a) Snapshots of $v\xc(x,t)$ and $n(x,t)$ for the field-free dynamics of the superposition state of the 1D He atom, over a half-Rabi cycle, whose dipole appears in Fig.~\ref{fig:1Dexamples}a. The adiabatically-exact and ALDA potentials lack the prominent step and peak structures of the exact. (b) Snapshots of the density and potentials for the e-H scattering problem of Fig.~\ref{fig:1Dexamples}b. The top panel shows the density at a time when the electron begins to reflect from the atom and partially transmit to the left, with the grey being the initial density for reference; the exact (black, left panel),  ALDA (red, right panel), $v\xc^{\rm S}$ (blue, right panel). The middle panels show the exact $v\xc(x,t)$ at that time (black), ALDA (red), $v\xc^{\rm S}$ (blue), and grey shows the exact $v\xc(x,0)$. The lower left shows the adiabatically-exact approximation (orange) for the exact $v\xc^{\rm W}$ component (black), and lower right the $v\c^{\rm T}$ component. 
Reproduced from Ref.~\cite{LM20b} with permission from the Royal Society of Chemistry. c) The exact and adiabatically-exact correlation potentials at the final time when the charge-transfer state is reached, for the resonantly-driven charge-transfer dynamics of Fig.~\ref{fig:1Dexamples}c. 
}
  \label{Steps}
\end{figure}

One common thread across all the examples in Sec.~\ref{sec:memory}, is the presence of prominent step and peak features in the exact xc potential, which are significant on the scale of the total KS potential. Figure~\ref{Steps} shows the exact xc potential in comparison with approximations including the adiabatically-exact, for those examples. 
The exact xc potential was computed by numerically inverting the propagation operator viewed as a function of the potential to target the exact density~\cite{FLNM18,LM18}. In the case of a  Slater determinant formed with a doubly-occupied orbital, the KS orbital is directly related to the exact density, and then the xc potential can alternatively be computed using analytical formulas~\cite{EFRM12} (see also Sec.~\ref{sec:He}).

A salient feature common to all the exact xc potentials over the range of different dynamics are step- and peak-like structure which are absent in  the adiabatic approximations, including the adiabatically-exact (although partially reproduced in case c)). 
These features are therefore non-adiabatic and their absence in adiabatic approximations yield inaccurate density-dynamics.

Fig~\ref{Steps}a shows different time-snapshots of the exact, adiabatically-exact and ALDA xc potentials for the 1D He atom superposition state of Fig~\ref{fig:1Dexamples}(a). Given that in this case the KS state is a doubly-occupied orbital, $v\x = -v\H/2$ has a simple well structure that smoothly cradles the density (the exact $n(\br,t) $ is shown as blue dotted), the step and peak are features of the correlation potential which often dominates the KS potential~\cite{LFSEM14,FNRM16,FLNM18,LM18,LM20b}. In the case shown, the KS state has a fundamentally different structure to the true state; instead, if the KS state is chosen with the same configuration of the true state, the step features are smaller but still appear, although their impact on the ensuing dynamics is less~\cite{LM20b}. 

In Fig.~\ref{Steps}b we display the xc potential and density for the scattering example of Fig~\ref{fig:1Dexamples}b, for the case when $\Phi(0) = \Psi(0)$~\cite{SLWM17,LSWM18}. The middle panel shows again a prominent peak and step feature  just to the right of $x= -10$a.u. at the shoulder of the electron density (top panel), which persists over time and contains the effective correlation needed for reflecting part of the density of the two KS electrons. Adiabatic approximations miss this structure and fail to reflect: the top right and middle panels show the density and xc potential for the ALDA  in red. 
The feature appears in the $v\c^{\rm T}$ component of the exact potential, and the lower panels show that while the adiabatically-exact approximation accurately captures $v\xc^{\rm W}$, it hardly resembles $v\c^{\rm T}$ at all. The figure also plots a non-adiabatic approximation,  $v\xc^{\rm S}$, which replaces the exact xc hole and one-body density-matrix in Eqs.~(\ref{eq:vxcW})--(\ref{eq:vcT}) with their KS counterparts; this yields a reasonable approximation to $v\xc^{\rm W}$ but gives zero for $v\c^{\rm T}$.

Finally, turning to the example of resonantly-driven charge-transfer of 
Fig.~\ref{fig:1Dexamples}c, we plot in Fig.~\ref{Steps}c the exact correlation potential at the time the charge-transfer state is reached. A step appears in the adiabatically-exact potential, however it has the wrong height. Ref.~\cite{FERM13} shows that the heights of the two steps in the limit of large separation between the donor (D) and acceptor (A) can be given by the differences in ionization potential and electron affinity indicated in the figure;  the adiabatically-exact step has a size equal to the $(N_D - 1)$-electron derivative discontinuity, and is smaller than that of the exact. As this step builds up over time during the charge transfer, there is also an oscillatory dynamical step similar in nature to that in part (a).

Thus these non-adiabatic step and peak features have been demonstrated on a wide range of dynamics, and we now briefly mention some beyond those above. Ref.~\cite{RG12} found them in the exact KS potential for a model describing the propagation of a single electron through an infinite semiconductor wire, using a quasi-particle wavepacket of nonzero crystal momentum added to the ground-state of a semiconductor. Refs.~\cite{HRCLG13,HRDG14} studied field-induced tunneling in a system of two or three spinless electrons in 1D.   Ref.~\cite{K16} showed barrier structures that were essential for autoionization processes in a 1D He atom.
Refs.~\cite{V08,KPV11,VKFAB11,MRHG14,FM14,FM14b} considered Hubbard models, while Ref.~\cite{DSH18} demonstrated their importance for single-electron transport through a quantum dot using  an Anderson model. Although computationally more convenient to demonstrate on  1D model systems, in Section~\ref{sec:He} we use the example of the real (three-dimensional) Helium atom to illustrate how the dynamical non-adiabatic features in the exact $v\xc$ shown in the 1D helium atom persist just as strongly in the three-dimensional  case.

Step and peak features are no strangers to DFT.
In the ground-state, they appear in situations associated with ``fractional charge" regions, for example when a molecule is near a metal surface~\cite{PPLB82,P85}, related to the derivative-discontinuity. They appear in interatomic regions of dissociating molecules~\cite{P85,AB85,BBS89,GB96,TMM09,HTR09,HRG16,KPS16,GG20,GVG18,KKS21,M22,GNBG22} where they are signatures of static correlation. Without the interatomic step, which is equal to the ionization potential difference between the two fragments, the KS system dissociates to unphysical fractional charges. Associated with the onset of the interatomic step is a peak, and recently a secondary peak was found to appear in the very low density region far to the side of the molecule that heralds the descent of the step back to zero asymptotically~\cite{GVG18,GNBG22}.
These step and peak features appear in the correlation potential and can be analyzed in terms of changes in the conditional amplitudes of the interacting and KS systems, defined through a factorization, $\Psi(\br_1,\br_2...\br_N) = \sqrt{\frac{n(\br_1)}{N}}\Psi_{\rm cond}(\br_2...\br_N)$ ~\cite{BBS89,GB96,TMM09,GVG18,GNBG22}. Although the exact exchange potential displays step features when the orbital dominating the density switches to one with a different asymptotic decay~\cite{LGB95,GKKG00,MKK11}, it does not capture the interatomic step in dissociating diatomic molecules which is a strong correlation effect~\cite{M17,TMM09,M22}.
  In the absence of nodal planes of the highest occupied molecular orbital (HOMO), the steps go back down to zero asymptotically, which is a striking difference with the non-adiabatic steps in the time-dependent case. 
  Even when there are nodal planes, steps, peaks, and diverging behavior appear as one traverses across the nodal plane of the HOMO in cases where the density has isotropic decay away from the system~\cite{DG02c,GGB16}. 
We stress here that the exact ground-state xc potential captures these features but does not capture the non-adiabatic steps in non-equilibrium TDDFT. 

In the linear response regime, steps have appeared that counter the electric field across long-range molecules~\cite{GSGB99}, and in the xc kernel near charge-transfer excitations associated with a derivative-discontinuity~\cite{HG12,HG13}.  Steps were found to be essential in Coulomb blockade phenomena, again related to fractional charges and the derivative-discontinuity~\cite{KSKV10}.
These cases involve perturbations around the ground-state, and the steps are associated with fractional charge effects. Beyond the response regime, dynamical steps were found in ionization processes, where an adiabatic approximation that depended on the fraction of charge remaining locally near the parent atom was able to approximate them~\cite{TGK08,LK05}. 

The non-adiabatic steps we are discussing here are a distinct phenomenon to all these cases: they are not related to fractional charges, ionization, nor is an external field necessary, and they are missing from any adiabatic approximation, that is, they cannot be captured by using any ground-state approximation. They appear in the kinetic component of the xc potential, Eq.~(\ref{eq:vcT}), and are instead related to the local KS velocities and accelerations in the system, as we will discuss next. 

\subsection{Relationship to local velocity and acceleration}
Earlier analysis on the two-electron systems have shown that the dynamical steps tend to be associated with the spatial integral of the local KS acceleration in the system~\cite{RG12,HRCLG13,EFRM12,FERM13,LFSEM14,M16}, while the peak structures are associated with a maximum in the local KS velocity, or at maximum curvature in the density in regions of low density. To see this, consider first the two-electron  spin-singlet where the KS initial state is chosen to be a Slater determinant. This means that  we always have a single spatial KS orbital that is doubly-occupied, which must have the form:
\begin{equation}\label{eqn:phi}
\varphi(\br,t) = \sqrt{n(\br,t)/2} e^{i\alpha(\br,t)}
\end{equation}
to reproduce the exact interacting density  with the phase~$\alpha(\br,t)$ related to the current-density~$\bj(\br,t)$ through the equation of continuity, 
\begin{equation}\label{eq:continuity}
   \nabla\cdot \bj =   \nabla\cdot\left(n(\br,t)\nabla \alpha(\br,t)\right)= - \frac{\partial}{\partial {t}}n(\br,t)\,.
   \end{equation}
  We note that the KS current-density $\bj\s(\br,t) = n\nabla\alpha(\br,t)$ may differ from the true current-density $\bj(\br,t)$ by a rotational component~\cite{AV05,MBAG02,SK16,TK09b,GM12,DLFM21} (although in 1D they are identical, $\bj\s(\br,t) =\bj(\br,t)$).
Inverting  Eq.~(\ref{KS_equation}) yields the exact KS potential:
\bea
\label{eq:vs}
    v\s(\br,t)&=&\frac{\nabla^2\sqrt{n(\br,t)}}{2\sqrt{n(\br,t)}}-\frac{\vert\nabla\alpha(\br,t)\vert^{2}}{2}-\frac{\partial\alpha(\br,t)}{\partial t}\\
    \nonumber
    &=&\frac{\nabla^2\sqrt{n(\br,t)}}{2\sqrt{n(\br,t)}}-\frac{1}{2} u^2(\br,t) - \int^\br_{\br_0} \frac{\partial{\bf u}({\bf{s'}},t)}{\partial t}\cdot\hat{{\bf s'}} ds'
\eea
where the second line is written in terms of local KS velocities and accelerations and is defined up to a global spatial constant. 
The second term in Eq.~(\ref{eq:vs}) is directly related to the local KS velocity, ${\bf u}(\br, t) = \frac{\bj\s(\br,t)}{n(\br,t)}$, and this term tends to produce peak structures, especially near density minima. 
The third term is related to a spatial-integral of the local KS acceleration, and relative to some reference position $\br_0$, it is given by
$\dot\alpha(\br,t) -\dot\alpha(\br_0,t)  \ = \int^\br_{\br_0} \partial_t{\bf u}({\bf s'},t)\cdot\hat{{\bf s}}' ds'$ where ${\bf s'}$ represents a path from $\br_0$ to $\br$. This shows that when there is a localized peak in the KS acceleration this term develops a step-like feature there. It should however be noted that the first term of Eq.~(\ref{eq:vs}) may also have structures that one can associate with local peaks and local steps:  these can partially cancel or enhance the velocity and acceleration terms. 
For example, in the trivial one-electron case one may get peaks in the local velocity and acceleration but $v\xc$ is zero, since the local curvature of the density can also display local steps and peaks especially near density-minima.
But steps that straddle across one side of a localized density to the other, leading to different asymptotic values of the potential in different directions, have a height given by the spatial integral of the localized KS acceleration across the system. We will show an example of this Sec.~\ref{sec:He} for the (three-dimensional) He atom. 

Eq.~(\ref{eq:vs}) holds for the special case of two-electron singlet systems where the KS initial state is chosen to be a Slater determinant. For a general choice and for $N$-electrons, the same expression holds but where the density, velocity, and acceleration are instead those of one of the occupied orbitals. Different orbitals yield different terms individually but the sum of the terms must be the same for each orbital; an example can be found in Ref.~\cite{M16}. 

\subsection{Approximations for the dynamical steps and peaks: Importance of $v\xc^{\rm T}$}

Although it is tempting to build an approximation directly from Eq.~(\ref{eq:vs}), it is not clear how to, because hiding in the expression is the external potential, whose functional dependence would somehow need to be subtracted out. In practise, the external potential is input as the potential function ({\it not} density functional) that is physically applied to the system, and only the xc and Hartree potentials are considered as functionals of the density and initial states. In fact this is essential, since without explicit input of the external potential one could not propagate with Eq.~(\ref{eq:vs}): it defines the physics of the problem being solved, and if instead it was represented by a density functional, causality and predictivity issues arise~\cite{MLB08,TU21}. We want an approximation for only the xc potential, not for the entire KS potential. Still,  it is interesting to note that an approximation for the non-adiabatic part of the full KS potential itself has been explored~\cite{EG20}, by transforming to the local instantaneous rest frame of the density, taking the adiabatic approximation to be exact in such a frame, and transforming back. This results in an approximation that adds precisely the last term of Eq.~(\ref{eq:vs}) to an adiabatic approximation. 
To finally hammer in the point of this paragraph, one can explicitly show that for two electrons in a single Slater determinant in one-dimension, this term is
\ben
-\frac{\partial}{\partial t}\int^x \frac{j(x',t)}{n(x',t)} dx'= v\ext(x,t) - \int^x \frac{\frac{\partial j^2(x',t)}{\partial x'} }{2n^2(x',t)} +
\frac{{\cal Q}(x',t)}{n(x',t)} dx'
\een
where ${\cal Q}(x',t) = {\cal D}_{x',x_2} \rho_1(x',x_2,t)\vert_{x_2 =x'} + \int \frac{\partial}{\partial x'}w(x',x_2)\rho_2(x',x_2,x',x_2,t)dx_2 $. 

Instead, Eqs.~(\ref{eq:vxcW})-(\ref{eq:vcT}) can be used to develop approximations. Moreover, analyses based on this decomposition have revealed that the non-adiabatic dynamical features figure more prominently in the kinetic term as compared to the interaction term. As  demonstrated in one of the examples in Fig.~\ref{Steps}, the step structures are a feature of the kinetic component $v\c^{\rm T}$ and are absent in adiabatic approximations to this term~\cite{FNRM16,FLNM18,LM18,LM20b,SLWM17,LSWM18}.
While adiabatic approximations may tend to approximate $v\xc^{\rm W}$ well, they do a poor job of capturing the structure in $v\c^{\rm T}$.
This is not surprising given the form of the two terms: While  the interaction term benefits from smoothing through the integral, the  appearance of multiple gradients in the kinetic component can create large step and peak features that evolve in time, especially in regions where the density has local minima. 

One such class of approximations is the density-matrix coupled approximation of Refs.~\cite{LM18,LM20b}. Here, the correlated one-body density matrix needed in $v\c^{\rm T}$ is replaced by an approximated one computed from the first equation of the BBGKY hierarchy, and propagated alongside the KS calculation. In the BBGKY equation the two-body density matrix is approximated by that of the KS system. 
This ensures that the diagonal part of this density matrix stays equal to the density of the KS system at all times, while the off-diagonal terms can differ from the KS one-body density matrix (as would be the case in an exact calculation).

Even though this approach satisfies most exact conditions known in TDDFT, and is able to capture the elusive dynamical steps and peaks,  it remains numerically too unstable to be of practical use. An improved approach within this class of approximations would include an approximation for two-body density matrix as a time-dependent functional of either or both the KS and correlated one-body density matrices.

Eq.~(\ref{eq:vcT}) suggests that the steps are a probe of the difference between the local character of the true and KS density-matrices near the diagonal (Eq.~\ref{eq:vcT}). Choosing initial KS states to minimize this difference may improve the performance of the adiabatic approximations at short times~\cite{FNRM16,LM20b}. The idea that the performance of an adiabatic approximation is connected with how small the variation of  natural orbital occupation numbers is was investigated in Ref.~\cite{LM20b}.  Ref.~\cite{LFSEM14} found that the largest step structures appeared at local minima of the largest occupation numbers. However it should also be noted that the impact of a large localized dynamical step or peak on the dynamics may be relatively small if the structure oscillates rapidly in time or is very localized in a region of small density~\cite{FLNM18,LM20b}.

\section{Case study: Dynamics in the Helium atom}
 \label{sec:He}
The examples illustrated in the previous section involved 1D systems, and a question then arises: are the non-adiabatic steps and peaks equally prominent in real three-dimensional systems? It is often said that correlation  effects are enhanced in reduced dimensionality~\cite{Brus2014, Ugeda2014}. 
A sensible comparison would be to compare the steps and peaks in a three-dimensional system undergoing analogous dynamics to one of the 1D examples. Thus here we study the
 field-free evolution of a 50:50 superposition state of the ground and first-excited dipole-accessible state of the real Helium atom~\cite{DLFM21}, and compare with the example (a) of Fig.~\ref{fig:1Dexamples} and Fig.~\ref{Steps}. 
 
 We find the exact interacting state at time $t$ in terms of the eigenstates of the He atom:   $1^1S_0$, denoted $\Psi_0$, and  singlet first excited state $2^1P_1$ that has angular quantum numbers $L=1$ and $M=0$, denoted here $\Psi_1$. Thus, 
\ben
    \vert\Psi(t)\rangle=\frac{1}{\sqrt{1+|a|^2}}\left(\vert \Psi_0 \rangle+a e^{-i\omega t}\vert \Psi_1\rangle\right)
    \label{eq:intstate}
\een 
where $\omega=E_{2^1P}- E_{1^1S} = 0.77980$ a.u.,  is the frequency with which the system oscillates. The parameter $a$ allows us to tune the proportion of the excited state $\Psi_1$ that goes into the superposition. For the analog to the 1D He example in Figs.~\ref{fig:1Dexamples}a and \ref{Steps}a, $a = 1$, the 50:50 superposition.  
In exact TDDFT, the density of this interacting state as a function of time
\ben
n(\br, t) = \frac{1}{1+\vert a\vert^2}\left(n_{0}(\br) + \vert a\vert^2 n_{1} (\br) + 2a n_{01}(\br)\cos(\omega t) \right)
\label{eq:dens}
\een
where $n_{q}(\br)  = 2\int \vert \Psi_q(\br, \br_2)\vert^2 d^3\br_2, q = 0, 1 $ and $n_{01}(\br) = 2\int \Psi_0(\br, \br_2) \Psi_1(\br, \br_2) d^3\br_2$, 
 is exactly reproduced by the non-interacting KS electrons. 
 
We note that although this may seem to be a very special case of dynamics, the results for the xc potential at any given time $t$ in fact apply to a far wider class of dynamics: due to Eq.~(\ref{eq:mem_cond}) the xc potentials we will find apply in any situation where the instantaneous true state is given by Eq.~\ref{eq:intstate} and an initial Slater determinant is chosen for the KS state. This means that independently of how the system reached instantaneous state at some time $t$ the xc potential at that time is the same.

To find the exact xc potential, we will need to invert the time-dependent KS equation for the three-dimensional case,  but first we need to find the exact densities of the interacting eigenstates and the transition density in order to construct the target density Eq.~(\ref{eq:dens}). 
We use the results of Refs.~\cite{Feistthesis,Feist08} to find the exact density. 
There, the system is solved using the close-coupling method, in which the angular part of the exact wavefunction is written in terms of coupled spherical harmonics while the radial part is obtained by the finite element discrete variable representation method~\cite{Feistthesis,Feist08,DLFM21}.



\subsection{Extracting the exact xc potential}
Here we choose the initial KS state as a Slater determinant: this is the natural choice if the state Eq.~(\ref{eq:intstate}) is reached from applying an external field to a ground-state and then turning the field off. One would use ground-state DFT to find the initial KS orbitals, and by the ground-state theorems, this is a Slater determinant. Since the KS evolution involves a one-body Hamiltonian, the state remains a single Slater determinant. 


This choice of initial KS state is the most relevant choice for simulations of general dynamics of the atom when beginning in the ground state. This is often the case for simulating experiments involving laser-driven dynamics, for example, or in photovoltaics where one models the initial photo-excitation process itself; and, as mentioned above, the xc potential we find at time $t$ is the same for any situation in which the true state Eq.~\ref{eq:intstate} is reached and a KS Slater determinant is chosen. 
  For the case of the He atom, the KS state is then a doubly-occupied orbital. 
The exact xc potential is then obtained from subtracting $v\ext = -2/|\br - \br'|$ and $v\H(\br,t)$ from Eq.~\ref{Vs_equation}. 
Further, one can isolate the correlation component by noting that for this KS state, $v\x(\br,t) = -v\H(\br,t)/2$. 

Thus, finding the exact xc  potential reduces to solving Eq.~(\ref{eq:continuity}) for $\alpha(\br, t)$ and inserting it into Eq.~(\ref{eq:vs}). We note that for a different choice of initial KS state, e.g. using a two-configuration state that is more similar to that of the actual interacting state, the inversion to find $v\xc$ involves an iterative numerical procedure~\cite{RL11,NRL13,RPL15}; some examples for the 1D analog of the He dynamics can be found in Refs.~\cite{FNRM16,FLNM18,LM18} (see also discussion in Sec.~\ref{sec:steps}). This could be a more natural state to begin the KS calculation in some situations, e.g.  if the state was  prepared in such a superposition at the initial time. 
The importance of judiciously choosing the KS initial state when using an adiabatic approximation has been realized and exploited in strong-field charge-migration simulations~\cite{BHMALGSL17,FBMHH21}.

\subsection{ Numerical Details }
We observe that the second equality in Eq.~(\ref{eq:continuity}) 
has the form of a Sturm-Liouville type equation, which yields a unique solution for $\alpha(\br, t)$ for a given boundary condition. Furthermore, thanks to the azimuthal symmetry of our density ($M=0$ at all times), we need solve this in effectively two dimensions, $(r,\theta)$.
We construct the explicit matrix representation of the operator $ \nabla\cdot\left(n(\br,t)\nabla \right)$ using the fourth order finite-difference scheme and subject to the boundary conditions,
\begin{equation}\label{boundary conditions}
\alpha(\br\to\infty,t)=0  \;\;\;\;\text{and}\;\;\;\;
\frac{\partial}{\partial{\theta}}\alpha(\br,t)\vert_{\theta=\pi,0}=0\,.
\end{equation}

We use a rectangular computational domain for the grid ($r,\theta$), extending from $0\to R=30$ a.u. in $r$ (the density is negligible this far from the nucleus) and $0\to \pi$ in $\theta$, and the boundary conditions, Eq. (\ref{boundary conditions}) translate to 
\begin{eqnarray}
\nonumber
\alpha(r=R,\theta,\varphi,t)&=&0\\ 
\frac{\partial}{\partial{\theta}}\alpha(r,\theta,\varphi, t)\vert_{\theta=\pi}&=&0\;\;\;\;\;\;
\frac{\partial}{\partial{\theta}}\alpha(r,\theta,\varphi, t)\vert_{\theta=0}=0
\end{eqnarray}
With the finite-difference scheme the matrix representation of the derivative operator  is highly sparse, and consequently the semi-local operator $ \nabla\cdot\left(n(\br,t)\nabla \right)$ is also, coupling only a few nearby grid points.  The dimensions of the matrix are $(N_{\theta} N_r) \times (N_{\theta} N_r)$ where $N_{\theta}=50$ and $N_{r}=301$ are the number of grid points in the ($\theta,r$) computational domain. 
The high sparsity also allows for efficient matrix inversion.  Despite the computational efficiency, caution is required to avoid numerical inaccuracies especially where the density becomes small. To ensure that our conclusions are robust, we restrict our analysis to regions where the inversion is fairly accurate. This is ensured by checking that the action of the matrix representing 
$\nabla\cdot  n(\br,t)\nabla$ on the solution vector $\alpha(\br, t)$ agrees with the right-hand-side of Eq.~(\ref{eq:continuity}).

In addition to the uniqueness entering the solution of Eq.~(\ref{eq:continuity}) as stated before, choosing the initial condition as $\alpha(\br,0) = 0$ fixes our initial state as $\phi(\br,0) = \sqrt{n(\br,0)/2}$. The Runge-Gross theorem then ensures that there is a unique $v\xc(\br,t)$ that reproduces the exact $n(\br, t)$ and yields a unique $\alpha(\br,t)$ at later times. It is not {\it a priori} obvious that the unique solution to Eq.~(\ref{eq:continuity})  with the boundary-condition Eq.~(\ref{boundary conditions}) applied with time $t$ as a parameter is compatible with the TDKS evolution, but the results do evolve smoothly in time. 
Moreover, the rapid decay of the density at large $r$ in the left-hand side  of Eq.(\ref{eq:continuity}) implies that this leads to the matrix elements corresponding to these points being killed off which renders the boundary conditions at $r=R$ superfluous. 


The numerical inversion of the matrix operator $\nabla\cdot n(\br,t)\nabla$  subject to the boundary conditions Eq.~\ref{boundary conditions} described above, produces the solution of Eq.~(\ref{eq:continuity}) for $\alpha(\br,t)$.  When used in  Eq.~(\ref{eq:vs}) this yields the KS potential,  $v\s(\br,t)$. 
We then calculate the exact Hartree potential $v\H(\br, t)$, by numerically inverting
\begin{equation}
\nabla^2 v\H(\br,t)=-4\pi n(\br,t)\,,
\end{equation} 
in a similar way, so that  $v\H$ and $v\ext$ can be subtracted from $v\s$ to find the xc potential. 

\subsection{Results}
The problem possesses several symmetry features which we leverage to simplify the analysis. The azimuthal symmetry mentioned earlier  and the fact that the states that enter the superposition are those for which $L=0$ and $L=1$, imply that the density, current, and potentials in the lower half-plane ($\pi/2 <\theta <\pi)$ evolve in exactly the same way as those in the upper half-plane ($0 <\theta <\pi/2)$ but a half-cycle out of phase, i.e. if $O(r,\theta,t)$ represents the above quantities, then  $O(r, \pi-\theta, t) = O(r, \theta, t+T/2)$. Another consequence of the simple form of the superposition is the fact $O(\br, T-t ) = O(\br, t)$.
It is therefore sufficient to look at time-snapshots only over a half cycle in one of the octants; the discussion will carry over to the other octants in accordance with the above conditions.

\begin{figure}
\includegraphics[width=0.5\textwidth]{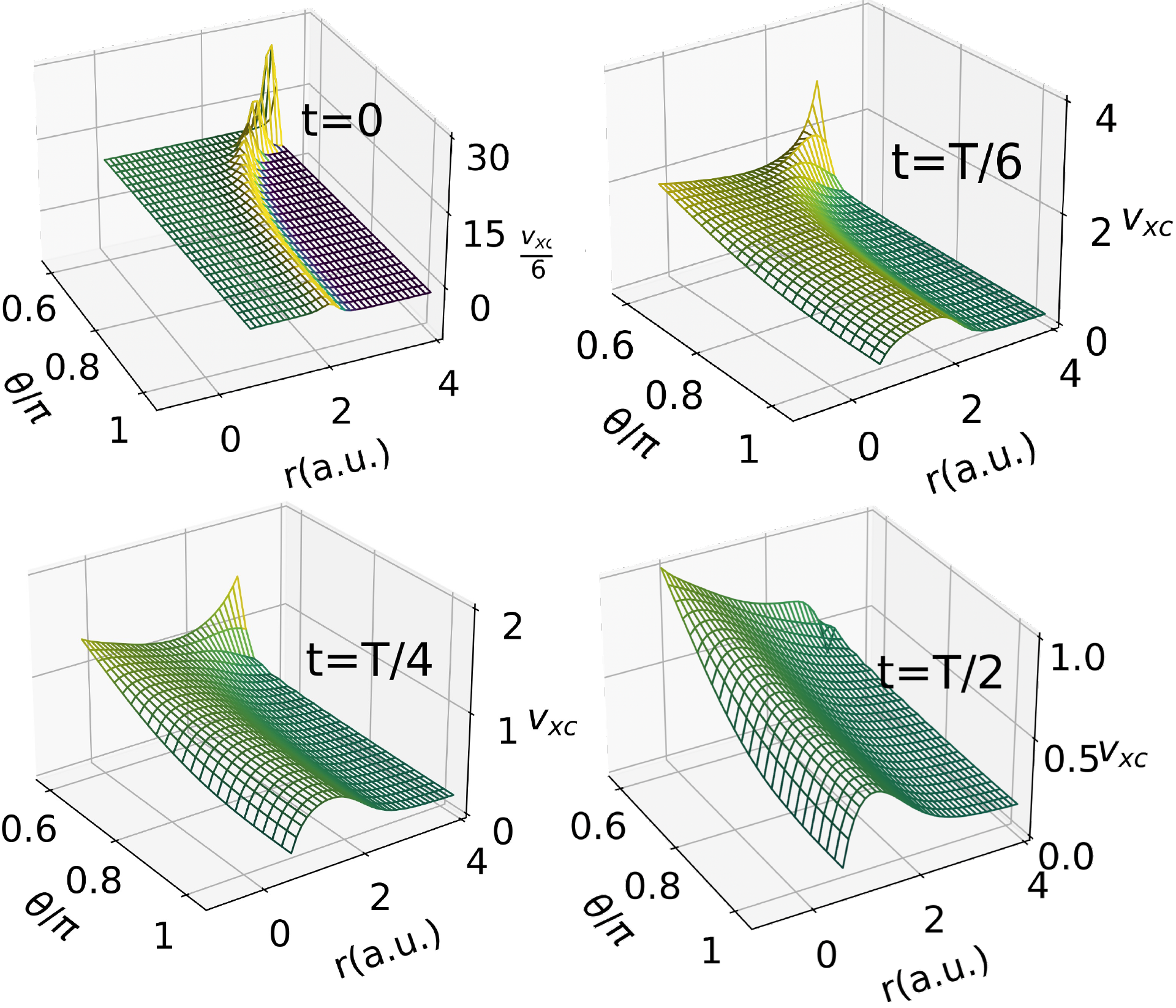} 
\caption{Exact xc potential, $v\xc(\br,t)$ 
for field-free evolution of the 50:50 superposition state ($a = 1$) of the He atom in the range $\pi/2 <\theta <\pi$ at times $t = 0, T/6, T/4$, and $T/2$}
\label{3Dvxc}
\end{figure}

In Figure~\ref{3Dvxc} we plot the snapshots of the xc potential, $v\xc({\br,t})$  at different fractions of the period of oscillation, $T = 2\pi/\omega = 8.057$ a.u.in the octant spanned by $[r =(0,4)]\times [\theta = (\pi/2,\pi)]$. 
We observe the xc-potential displays a prominent peak and a step across $r$ at $t=0$ in the region swept by $\pi/2 <\theta <\pi$ that decrease in magnitude over the first half-period time until vanishing from the octant and appearing on the other side of $\theta=\pi/2.$ 
These features have been shown to be completely missing in adiabatic approximations for 1D cases (e.g. Fig.~\ref{fig:1Dexamples}) yet they often dominate the KS potential. Here we find these non-adiabatic features persist just as prominently, in analogous dynamics of the real three-dimensional He atom~\cite{DLFM21}. This justifies that conclusions drawn from previous studies involving 1D systems do apply to real systems as well and that these strong correlation effects do not arise from dimensional reduction~\cite{Brus2014,Ugeda2014}. We expect that the lack of the step and peak features in adiabatic approximations lead to less structured density profiles, as was seen in 1D cases~\cite{LM20b}. 

 It is important to mention here that at any instant of times for our particular superposition state, the correlation potential goes to the same time-dependent constant asymptotically in every direction in the lower octant, and a different time-dependent constant in the upper octant. The flatness of the KS potential asymptotically in each octant is consistent with a zero current-density there.  Our inversion to find the potential very close to $\theta=\pi/2$ is not reliable near the density's local minimum along this line:  Looking across the $\theta-$direction, we observe a sharp step and peak in $v\c$ at $\theta = \pi/2$ at large $r$, which yields a force that prevents the KS current vectors from crossing the $xy$-plane, in a manner consistent with the behavior of the true current vectors~\cite{DLFM21}. These features make a numerical inversion near $\theta = \pi/2$ difficult. At $\theta = \pi/2$, the density of the $P$-state $\Psi_1$ vanishes, and the large and sharp change in the potential reminds us of the divergent behavior along the highest-occupied molecular orbital nodal plane found in the asymptotic region of the ground-state potential~\cite{GGB16} mentioned in Sec.~\ref{sec:steps} for cases where the density has the same decay in all directions. But in our case the density does decay differently along this plane than in other directions, and  it cannot be captured by any adiabatic approximation.  

As discussed in Sec.~\ref{sec:steps}, we find that a significant contribution to the  peak is related to the local KS velocity, while the step when a cut is taken across a fixed $\theta$ is related to the radial integral of the local acceleration, $\dot\alpha(r,t) = \int^r \partial_t{\bf u}(\br',t).\hat{\br}' dr'$, and likewise for the step at $\pi/2$ in the $\theta$-direction.  We will explicitly demonstrate this in Sec.~\ref{sec:non-intHe}.

\begin{figure}[H]
\includegraphics[width=0.5\textwidth]{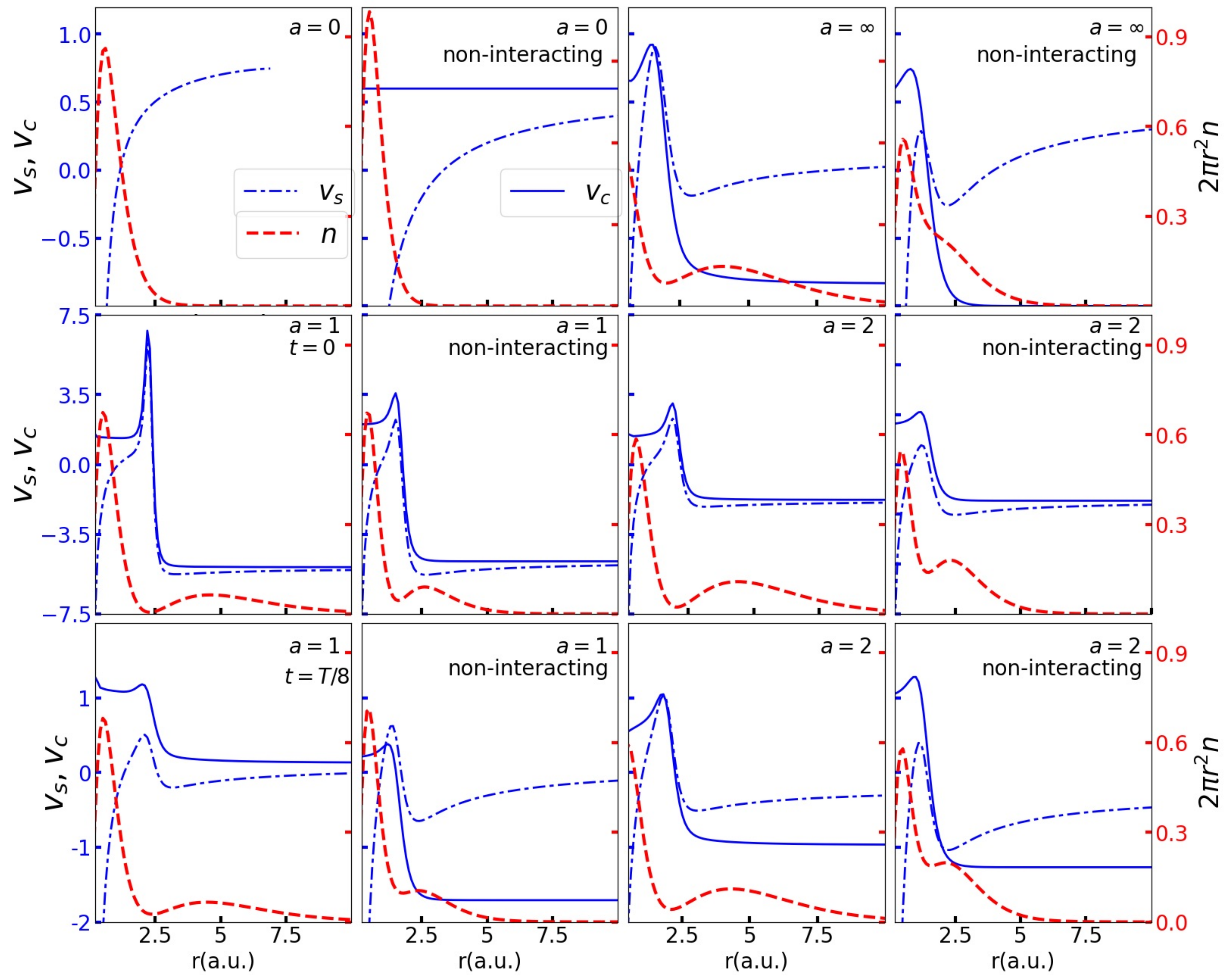}
\caption{Comparison between interacting and non-interacting helium atom. Columns 1 \& 3 correspond to the interacting case while columns 2 \& 4 correspond to the non-interacting case.
Top panel shows $n(\br,t)$ (dashed red), $v\s(\br,t)$ (dashed blue)  and correlation $v\c({\br,t})$ (solid blue) for purely ground state ($a =0$, columns 1 \& 2) and purely excited state ($a=\infty$, columns 3 \& 4), at a cross section taken at  $\theta = 0.75\pi$~\footnote{We have omitted the correlation potential for $a=0$ from the figure because it is so small that it falls within the numerical uncertainty of our inversion. Instead, it may be found in Ref.~\cite{UG94}. }. Middle panel: $n(\br,t)$, $v\s({\br,t})$ and $v\c({\br,t})$ plotted for superposition states 50:50 ($ a=1$, columns 1 \& 2) and 20:80 ($a=2$, columns 3 \& 4) at time $t=0$.
Lower panel: Same quantities as middle panel for $t=T/8$.  } 
\label{fig:vc_cmp}
\end{figure}

Taking different superpositions of the ground and excited states is further evidence that the step and peak features are universally present in real three-dimensional systems. The first and third columns of Figure~\ref{fig:vc_cmp} shows the KS and correlation potentials at the initial time, when $a$ in Eq.~\ref{eq:intstate} is changed through $0,1, 2,\infty$, going from a pure ground-state, through to the purely excited state; these two limits are of course time-independent. The correlation potential in the purely excited case (top panel, third row) has a barrier with a structure that is not dissimilar to the 1D case in magnitude and in shape (Figs. 3 and 4 in Ref.~\cite{EM12}); the KS potential must be such to maintain the constant excited $^1P$ density at all times with a non-interacting doubly-occupied orbital.    
(For all cases,  $v\s$ goes asymptotically like a constant $-1/r$; the different scales makes this behavior less apparent in some cases.)

\subsection{``Non-Interacting Helium"}
\label{sec:non-intHe}

As observed in Sec.~\ref{sec:steps}, it is the kinetic component $v\c^{\rm T}$ that is largely responsible for the  non-adiabatic step and peak features, and that even the best adiabatic approximation to this term, the adiabatically-exact,  does not capture them. 
 This component of the xc potential depends on spatial derivatives of the difference between the true and KS one-body density matrix (Eq.~(\ref{eq:vcT})), and so is significantly affected by the choice of initial KS state, as discussed earlier. In fact, even considering the (unphysical) limit of zero electron-interaction in the He atom, $v\xc = v\c^{\rm T}$ is non-zero when the initial KS state is chosen differently to that of this non-interacting atom. This raises the theoretical question of how do the steps and peaks compare in such a system to the physical situation, that is, how are these affected by electron-interaction itself, as opposed to configurational effects. 

To this end, we consider here a ``noninteracting He" atom, and prepare our system in a superposition of the ground state and an excited state such that the configuration is the same as that of the actual interacting  He atom considered in the previous subsection, except for the fact that the wavefunctions that go into the superposition are products of the ordinary hydrogenic wavefunctions.
That is, in Eq.~(\ref{eq:intstate}), instead of the interacting ground state, $\vert \Psi_0 \rangle=\vert 1^1S_0 \rangle$ and the first excited state $\vert \Psi_1 \rangle=\vert 2^1P_1 \rangle$, we prepare our system in the following superposition
\ben
    \vert\Psi^{(0)}(t)\rangle=\frac{1}{\sqrt{1+|a|^2}}\left(\vert \Psi^{(0)}_0 \rangle+a e^{-i\omega t}\vert \Psi^{(0)}_1\rangle\right)
    \label{eq:nonintstate}
\een 
where $\Psi^{(0)}_0(r_1,r_2)=\phi_1(r_1)\phi_1(r_2)$ and $\Psi^{(0)}_1=\frac{1}{\sqrt{2}}[\phi_1(r_1)\phi_2(r_2)+\phi_1(r_2)\phi_2(r_1)]$ are composed in terms of the products of hydrogenic wavefunctions $\phi_1$ and $\phi_2$
\begin{equation}
  \phi_1(r)=\sqrt{\frac{8}{\pi}}e^{-2r}
  ,\;\;\
    \phi_2(r)=\frac{\cos\theta}{\sqrt{\pi}} r e^{-r}
\end{equation}
Following the same steps as involved in the solution of Eq.~(\ref{eq:continuity}) we analyse the exact KS potential $v\s$ along with the exact correlation potential $v\c({\br,t})$ for this non-interacting version of the Helium atom. The second and fourth columns of Figure~(\ref{fig:vc_cmp}) present these potentials for $a = 0, 1, 2$ and $\infty$. Comparing with the first and third columns we see that they display structures that resemble the ones found in the interacting case. We find that the peaks and the steps follow a similar trend  to that displayed by the interacting system when the proportions of ground and excited states are changed. 

The appearance of such dominating steps in
the correlation potential here is fundamentally linked to the difference
in configurations of the interacting superposition state and KS Slater determinant rather than being a direct consequence of electron interaction. Tuning down the electron-interaction dampens the
peak but the step remains. Both these features have a role in   somehow nudging non-interacting electrons to be in separate lobes of the density, with the peak containing dynamical Coulomb interaction effects further enhancing the separation. An accurate approximation for the kinetic component $v\c^{\rm T}$ is required to capture these features. 

\begin{figure}[H]
\includegraphics[width=0.5\textwidth,height=0.35\textwidth]{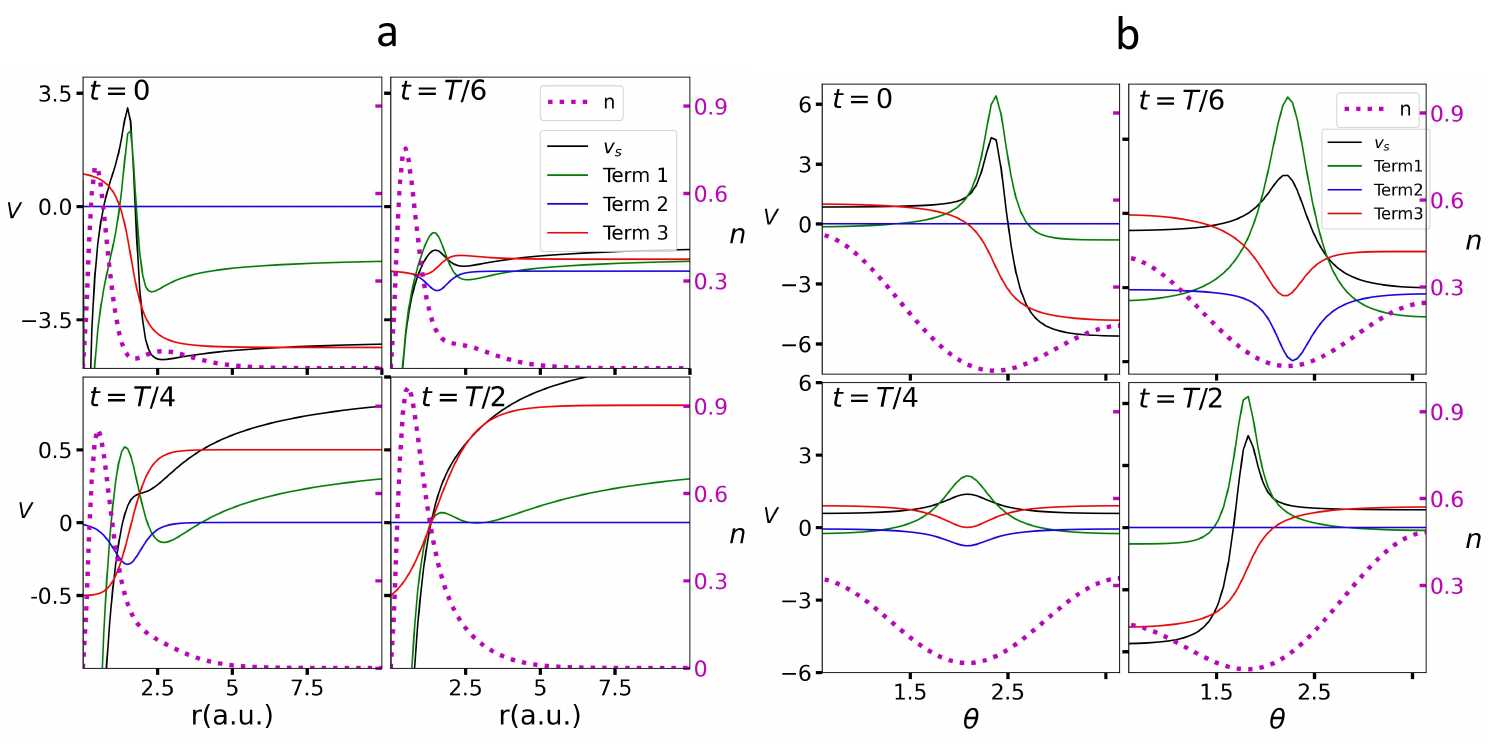}
\caption{(a) Decomposition of $v\s(\br,t)$: Terms 1-3 are the three terms in Eq.(\ref{eq:vs}) plotted as a function of $\theta$ at r=2.0 a.u. 
(b) Same three terms now plotted across as a function of $r$ at $\theta=0.75\pi$. 
In both (a) and (b), the density, $n(r,t)$ (dotted magenta) as a function of $\theta$ is plotted on a different scale at the same r. }
\label{fig:vs_cpts}
\end{figure}

Finally, we demonstrate with this non-interacting system, the decomposition of the exact KS potential in the second line of Eq.~(\ref{eq:vs}). In particular, 
the second term depends on the local velocity, and yields a peak structure, while the third term dependent on a spatial integral of the local acceleration is 
primarily responsible for the step. Having analytic expressions for the ``true" wavefunction gives analytic expressions for 
$n(\br,t)$ and $\bj(\br,t)$, and so allow  us calculate the local acceleration and velocity in a straightforward way. 
Figure~\ref{fig:vs_cpts} verifies the assertions above. We note that the first term in Eq.~(\ref{eq:vs}), that depends on the curvature of the instantaneous density, also can yield peaks near the minimum of the density; in fact at the initial time, where the current-density is zero, the peak structure arises predominantly from the first term. The first term can also yield localized step-like structures, but it is only the third term that can yield a step structure that asymptotes to different constants away from a localized density.
Interestingly, in this case, $\bj\s(\br,t) = \bj(\br,t)$, in contrast to the physical interacting He atom where snapshots of their difference can be found in Ref.~\cite{DLFM21}.


\section{Conclusions and Outlook}
\label{sec:concs}
TDDFT tasks the potential driving the non-interacting electronic system with reproducing the exact time-dependent density of interacting electrons at all times. As a result,  the exact time dependent xc potential must choreograph  a rather unusual dance through a landscape of dynamical steps and peaks that nudge non-interacting electrons to evolve with the same density as Coulomb-interacting ones. In this work, we have reviewed aspects of what is known about these steps and demonstrated them on a real three-dimensional system, the He atom. They are distinct from step and peak features that arise in the ground-state case which tend to be associated with fractional charge and static correlation issues. The dynamical steps and peaks instead appear generically in dynamics far from the ground-state, associated with a kinetic component to the xc potential that is sensitive to the difference of the exact and KS one-body density matrices near the diagonal. They have a non-local dependence on the density in time, and are completely absent in any adiabatic approximation. They also have a non-local dependence on the density in space, as can be seen e.g. from the different asymptotes of the potential. The entanglement of spatial and time non-locality in TDDFT has been pointed out in extended systems in linear response, where approximations that build in frequency-dependence while remaining local in space violate the zero-force theorem~\cite{VK96,TDDFTbook12,GVbook}, and here we see the time- and space- entanglement is a generic feature in non-perturbative dynamics of finite systems as well. The lack of these structures in approximations in use today lead to errors, including spurious peak-shifting that can muddle interpretation of pump-probe spectra.

Going beyond the adiabatic approximations, functionals that have explicit dependence on the instantaneous
orbitals incorporate memory (e.g. time-dependent
exact exchange~\cite{UGG95,LHRC17} but the orbital-dependence of these commonly
only involves exchange which is inadequate to capture these dynamical steps, as they appear in $v\c^{\rm T}$. Building practical non-adiabatic approximations for these features have so far proven elusive, although the recent density-matrix coupled approximations show it may be possible.  The hope is that such approximations would increase the reliability of real-time TDDFT for dynamics of electrons in non-perturbative fields. The challenge is certainly worth it. 

\acknowledgements{
Financial support from the National Science Foundation Award CHE-1940333 (DD)  and from the Department
of Energy, Office of Basic Energy Sciences, Division of Chemical
Sciences, Geosciences and Biosciences under Award No. DESC0020044 (NTM, LL), and the
 European Union's
Horizon 2020 research and innovation programme under the
Marie Sk{\l}odowska-Curie grant agreement No 101030447 (LL) are gratefully acknowledged. }
\bibliography{./ref.bib}

\end{document}